\documentclass[10pt,conference]{IEEEtran}
\usepackage{cite}
\usepackage{graphicx}
\usepackage{amsmath,amssymb,amsfonts}
\usepackage{algorithmic}
\usepackage{textcomp}
\usepackage{xcolor}
\usepackage{tikz}
\usetikzlibrary{positioning, arrows.meta, shapes.multipart, fit}

\usepackage[top=0.75in, bottom=1in, left=1in, right=1in]{geometry}
\IEEEoverridecommandlockouts

\begin{document}

\title{A4FN: an Agentic AI Architecture for Autonomous Flying Networks

\thanks{This work is financed by National Funds through the FCT -- Fundação para a Ciência e a Tecnologia, I.P. (Portuguese Foundation for Science and Technology) within the project FALCON, with reference 2023.15645.PEX (https://doi.org/10.54499/2023.15645.PEX).}
}

\author{
    \IEEEauthorblockN{André Coelho, Pedro Ribeiro, Helder Fontes, Rui Campos}
    \IEEEauthorblockA{
        INESC TEC, Faculdade de Engenharia, Universidade do Porto, Portugal \\
        Emails: \{andre.f.coelho, pedro.m.ribeiro, helder.m.fontes, rui.l.campos\}@inesctec.pt}
}

\maketitle

\begin{abstract}
This position paper presents A4FN, an Agentic Artificial Intelligence (AI) architecture for intent-driven automation in Flying Networks (FNs) using Unmanned Aerial Vehicles (UAVs) as access nodes. A4FN leverages Generative AI and Large Language Models (LLMs) to enable real-time, context-aware network control via a distributed agentic system. It comprises two components: the Perception Agent (PA), which semantically interprets multimodal input -- including imagery, audio, and telemetry data -- from UAV-mounted sensors to derive Service Level Specifications (SLSs); and the Decision-and-Action Agent (DAA), which reconfigures the network based on inferred intents. A4FN embodies key properties of Agentic AI, including autonomy, goal-driven reasoning, and continuous perception-action cycles. Designed for mission-critical, infrastructure-limited scenarios such as disaster response, it supports adaptive reconfiguration, dynamic resource management, and interoperability with emerging wireless technologies. The paper details the A4FN architecture, its core innovations, and open research challenges in multi-agent coordination and Agentic AI integration in next-generation FNs.
\end{abstract}

\begin{IEEEkeywords}
Large Language Models (LLMs), Autonomous Flying Networks, Agentic AI, Generative AI, Unmanned Aerial Vehicles (UAVs).
\end{IEEEkeywords}

\section{Introduction}
Wireless networks play a critical role in modern infrastructures, enabling use cases such as disaster response, emergency communications, environmental monitoring, and precision agriculture~\cite{Ribeiro2024}. These scenarios require connectivity solutions that are not only robust and scalable but also adaptive to highly dynamic and unpredictable conditions. Even though ground-based wireless systems are effective in static deployments, they face significant limitations in high-mobility or time-sensitive contexts due to restricted flexibility and limited deployment agility.

\begin{figure}
\centering 
\includegraphics[width=1\columnwidth]{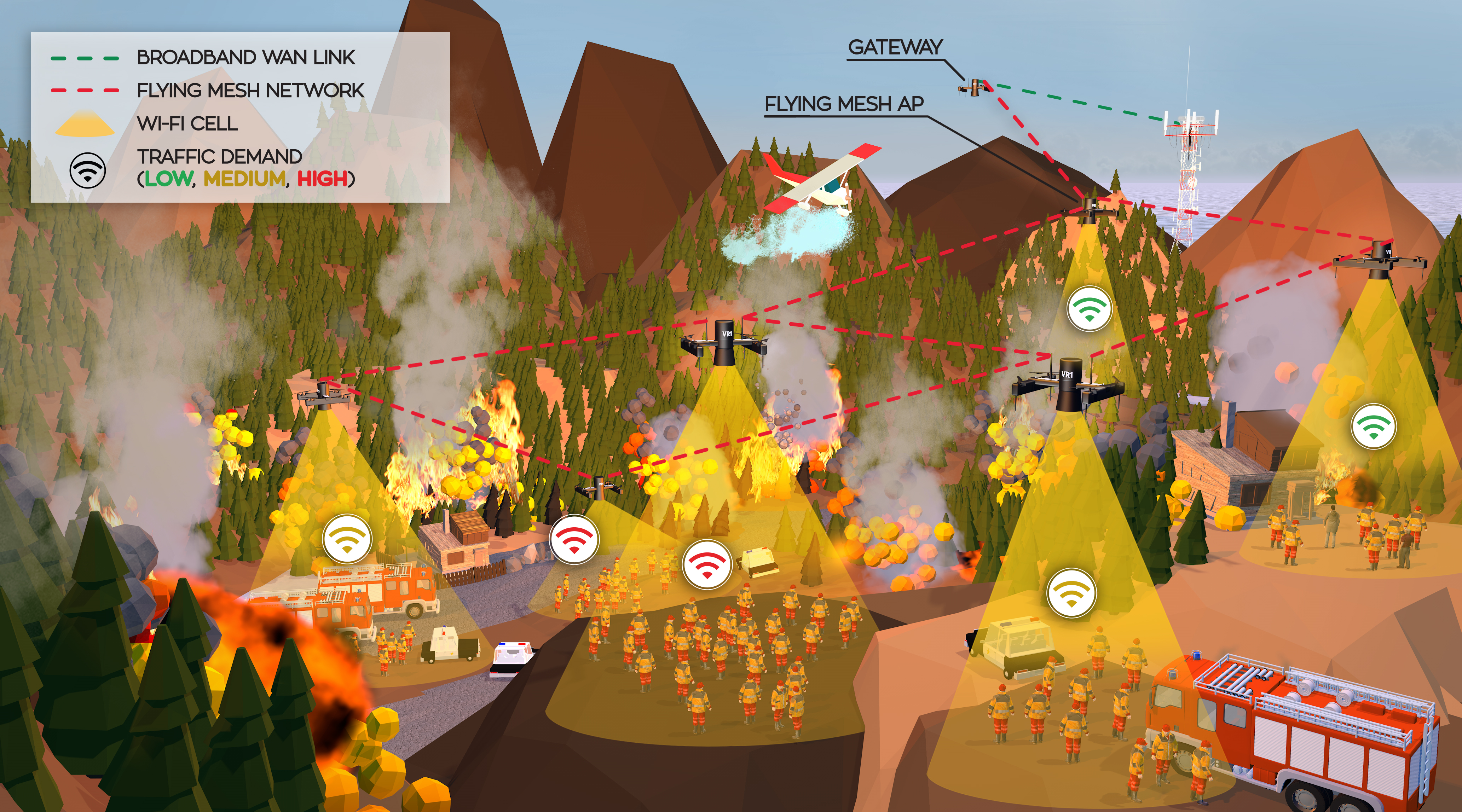}
\vspace*{-4mm}
\caption{Deployment of a flying network providing wireless connectivity in a disaster response scenario.} 
\label{fig:emergency-scenario} 
\end{figure}

Flying Networks (FNs), which leverage Unmanned Aerial Vehicles (UAVs) as mobile access points and base stations, have emerged as a promising alternative. They offer rapid deployment, dynamic coverage, and the ability to reach otherwise inaccessible areas ~\cite{Coelho2022}. Fig.~\ref{fig:emergency-scenario} illustrates a representative deployment of an FN in a disaster response scenario, where the mobility of UAVs enables wireless connectivity with dynamic reconfiguration capabilities. However, real-world FN deployments remain constrained by limited environmental awareness and continued reliance on static configurations and manual control. Most existing approaches assume prior knowledge of user locations and traffic demands -- assumptions that are often invalid in practical scenarios~\cite{Ribeiro2024, Wang2024_2}.

Recent progress in Artificial Intelligence (AI), particularly in Large Language Models (LLMs) such as GPT-4, Gemini, and LLaMA, has introduced new capabilities in semantic reasoning, intent interpretation, and multimodal perception. These developments create opportunities for a new generation of AI-native networking systems. Yet, their application in the autonomous management of FNs remains largely unexplored.

This position paper introduces the \textbf{Agentic AI Architecture for Autonomous Flying Networks (A4FN)}, a conceptual architecture for intent-driven automation in FNs using Unmanned Aerial Vehicle (UAV)-mounted sensing and communications capabilities. Intent-driven automation involves the automatic derivation of network requirements (intents) from high-level context, followed by corresponding network reconfigurations, eliminating the need for manual input or predefined rules. In contrast to traditional FN architectures that depend on static configurations, A4FN utilizes LLM-powered reasoning to infer user intent from multimodal sensor data.
A4FN integrates LLM-powered intelligence into the network control loop through two agentic components: the \textit{Perception Agent (PA)}, which semantically interprets multimodal input, including imagery, audio, and network telemetry data, to derive real-time Service Level Specifications (SLSs); and the \textit{Decision-and-Action Agent (DAA)}, which reconfigures the network based on the inferred intents.

A4FN embodies key properties of Agentic AI, including autonomy, goal-directed reasoning, and continuous perception-action cycles. Rather than conventional terrestrial networks, FNs must directly interact with their physical environment -- through perception enabled by UAV-mounted sensors and through actuation via UAV positioning and resource orchestration. These characteristics make FNs particularly suited for agentic architectures capable of interpreting complex environmental inputs and adapting system behavior accordingly. By embedding such intelligence into the control loop, A4FN contributes to the vision of Artificial General Intelligence (AGI)-native telecom infrastructure~\cite{Chaccour2024}, paving the way for resilient, autonomous, and context-aware wireless systems.

The remainder of this paper is structured as follows. Section~\ref{sec:related-work} reviews related work on FNs, LLMs, and intent-based networking. Section~\ref{sec:proposed-approach} details the proposed A4FN architecture. Section~\ref{sec:challenges} outlines the main challenges and research opportunities. Section~\ref{sec:conclusions} concludes the paper.

\section{Related Work \label{sec:related-work}}
FNs, primarily supported by UAVs, have emerged as a flexible and scalable solution for extending wireless network coverage in dynamic and infrastructure-limited environments. Extensive research has focused on optimizing UAV placement and resource allocation to improve network performance. For example, Almeida et al.~\cite{Almeida2018} proposed a traffic-aware, multi-tier deployment approach that enhances throughput and reduces latency by adapting UAV positions to traffic demands. Coelho et al.~\cite{Coelho2022} expanded on this work by introducing network slicing-aware algorithms that dynamically adjust UAV placement and allocate resources to meet heterogeneous service requirements.

More recently, the paradigm of intent-based networking has gained traction in wireless communications. Aklamanu et al.~\cite{Aklamanu2018} demonstrated the benefits of abstraction layers in simplifying real-time 5G service provisioning. Mehmood et al.~\cite{MEHMOOD2023} provided a comprehensive review of intent-based frameworks, highlighting their potential to enable autonomous and adaptive network management. However, these approaches often rely on static templates or domain-specific rule sets, which limit their responsiveness in real-world deployments.

To address the growing complexity and dynamism of modern networks, recent research has begun exploring Agentic AI -- an emerging paradigm characterized by autonomous, goal-directed systems capable of perception, reasoning, and action. Early explorations of this concept include the integration of Large Language Models (LLMs) into robotics and control systems. Vemprala et al.~\cite{Vemprala2024} demonstrated how LLMs such as ChatGPT can translate natural language instructions into executable actions in robotic platforms. Similarly, Tazir et al.~\cite{Tazir2023} used multimodal inputs for UAV control, emphasizing the role of semantic understanding in autonomous behavior.

The application of LLMs in FN control remains in its early stages, but recent efforts have shown promising directions. Prior works have explored LLM-based techniques for flight path generation, traffic prediction, channel modeling, and network simulation~\cite{Sun2024,Javaid2024,Javaid2024_2,Sun2024_2,Rong2024}. Notably, Sun et al.~\cite{Sun2024} proposed a two-stage optimization framework in which LLMs generate UAV flight trajectories and Graph Neural Networks perform resource allocation. Zheng et al.~\cite{Zheng2024} introduced a semantic communication system in which UAVs perform intent extraction and data delivery while preserving semantic fidelity. In~\cite{Nunes2025}, we introduced FLUC, a framework that uses LLMs to translate natural language commands into executable UAV flight plans; however, it operates under a prompt-driven paradigm, which depends on human input.

Security-focused applications have also emerged. Piggott et al.~\cite{Piggott2023} introduced Net-GPT, a fine-tuned LLM capable of executing protocol-compliant man-in-the-middle attacks in UAV-based networks. Li et al.~\cite{Li2024} presented LEDMA, an LLM-enabled multi-objective evolutionary algorithm for Integrated Sensing and Communications (ISAC), optimizing UAV placement, power allocation, and transmission strategies. Wang et al.~\cite{Wang2024_2} employed LLM-generated meta-prompts to optimize UAV positioning for both access and backhaul, demonstrating improved performance over traditional heuristics.

Despite these advancements, existing FN control architectures -- whether rule-based, heuristic, or reinforcement learning (RL)-based -- continue to lack the generalization, multimodal perception, and semantic interpretation capabilities required for fully autonomous operation. The integration of LLMs to enable real-time, intent-driven control in FNs remains an open and largely unexplored research direction, which this position paper aims to investigate and advance.

\section{Agentic AI Architecture for Autonomous Flying Networks\label{sec:proposed-approach}}

A4FN is a conceptual architecture designed to enable real-time, intent-driven control of FNs. Existing FN solutions often rely on static or semi-static configurations and require human intervention for reconfiguration, limiting responsiveness and scalability -- especially in dynamic, mission-critical scenarios such as disaster response.

A4FN addresses these limitations by embedding LLM-powered agentic components capable of: i) analyzing multimodal input; ii) deriving context-aware SLSs; and iii) reconfiguring the FN in response to evolving conditions. This architecture represents a shift from rule-based control toward semantic, goal-oriented management grounded in Agentic AI principles, enhancing adaptability, autonomy, and scalability in UAV-assisted wireless networks.

\subsection{Conceptual Architecture}
A4FN comprises two key agentic modules: the \textit{Perception Agent (PA)} and the \textit{Decision-and-Action Agent (DAA)}. Together, they form a distributed, closed-loop system capable of semantic interpretation, autonomous reasoning, and real-time reconfiguration.

\begin{figure}
\centering
\includegraphics[width=0.9\columnwidth]{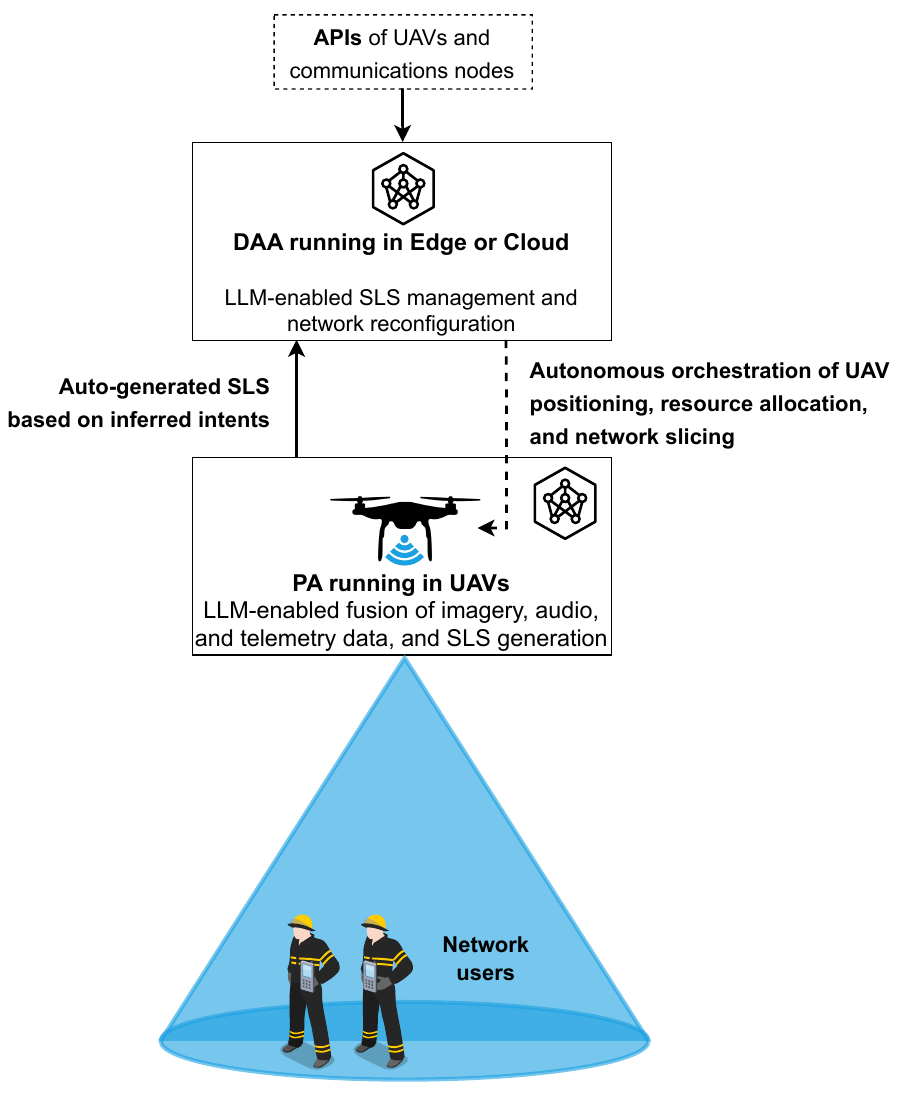}
\vspace*{-3mm}
\caption{Illustration of the A4FN conceptual architecture.}
\label{fig:A4FN-enhanced}
\end{figure}

The overall architecture is illustrated in Fig.~\ref{fig:A4FN-enhanced}. The PA, deployed onboard UAVs, acts as the perception module of the system. Built upon a multimodal LLM, it fuses diverse inputs, including imagery, audio, and telemetry, to infer the operational context of the FN. Through semantic reasoning, the PA derives dynamic SLSs that reflect user intent, mobility, traffic variations, and environmental conditions. These real-time SLSs are transmitted to the DAA, which is hosted at the edge or in the cloud and maintains a global view of the FN. The DAA uses this information to orchestrate UAV positioning, resource allocation, and network slicing. It interfaces with network elements via standard Application Programming Interfaces (APIs), enabling closed-loop, fully autonomous control.

As an illustrative scenario, consider a forest fire where UAVs equipped with thermal imaging cameras, microphones, and air-quality sensors detect rapidly rising temperatures, signs of smoke, and audio signals indicating distressed human voices. The PA, onboard UAVs, semantically interprets this multimodal data, triggering a critical-intent SLS specifying low-latency, high-reliability communications needs. This SLS is sent to the DAA at the edge or cloud, which autonomously repositions nearby UAVs, dynamically allocates spectrum resources, prioritizes network slices for emergency responders, and activates supplementary communications channels (e.g., Wi-Fi 8 mesh networks), enabling rapid, autonomous adaptation without human intervention.

\subsection{Key Innovations}

The A4FN approach introduces a set of innovations that address several core limitations of existing FN architectures. Central to this design is the use of \textbf{dynamic SLS} generation, enabling the PA to continuously interpret multimodal data and derive context-aware, intent-driven service definitions. This capability replaces static or rule-based configurations, allowing the system to respond in real time to user mobility, traffic fluctuations, and environmental conditions.

A4FN also advances \textbf{intent-driven reconfiguration} by leveraging the semantic reasoning capabilities of LLMs. Rather than depending on low-level control policies, the architecture abstracts operational decisions into high-level intents, thereby simplifying network orchestration and reducing reliance on manual intervention or expert tuning. This abstraction supports more intuitive and adaptive network management, including the potential for natural-language-driven control.

Additionally, A4FN enables \textbf{real-time UAV placement and resource allocation} through the integration of AI-driven or heuristic algorithms. These algorithms allow the DAA to autonomously adjust network topology, allocate communications resources, and manage slices to sustain service quality in dynamic environments. This on-demand adaptability reduces latency, improves energy-efficiency, and ensures operational continuity in mission-critical scenarios.

Finally, A4FN is designed for \textbf{compatibility with emerging wireless technologies}, including 6G and Wi-Fi 8. Its future-proof architecture supports seamless integration across a range of deployment contexts -- from disaster response to dense urban networks -- ensuring scalability, interoperability, and sustainable operation.

These innovations position A4FN as a foundational implementation of Agentic AI for autonomous and intent-aware network control.

\subsection{Potential applications}
A4FN is particularly suited for enhancing the resilience and agility of communications infrastructure in \textbf{infrastructure-limited, high-demand, or rapidly evolving environments}. In disaster response scenarios -- where ground infrastructure may be unavailable -- it offers a scalable, autonomous alternative to static, over-provisioned systems. Its agentic components can dynamically prioritize traffic for heterogeneous users, including civilians and emergency personnel, while optimizing resource use.

In urban contexts, A4FN supports rapid deployment during high-density events such as public festivals and sporting events. Its capacity to adaptively manage coverage and load reduces congestion, ensures fairness, and minimizes the energy and infrastructure footprint.

A4FN also holds potential for environmental monitoring and early-warning systems. It can autonomously provide communications support for IoT-based sensors (e.g., gas, temperature) and provide real-time video from UAVs for hazard detection. In post-disaster contexts, it serves as a self-healing network backbone, reestablishing connectivity and supporting coordinated response.

Its flexibility, scalability, and autonomy make A4FN applicable across a wide range of operational domains, from temporary coverage extension to long-term infrastructure augmentation, where intelligent, adaptive control is critical.

\subsection{Alignment with State-of-the-Art}

A4FN builds on recent advances in LLMs, intent-based networking, and autonomous multi-agent systems to overcome long-standing limitations in current FN architectures. Most existing approaches rely on static topologies, pre-defined input templates, or reinforcement learning methods with limited generalization. In contrast, A4FN introduces a dynamic, semantic, and multimodal alternative grounded in Agentic AI.

The architecture advances intent-based networking by generating and interpreting real-time SLSs from diverse sensor inputs. This marks a shift from rule-based templates to continuous, context-aware adaptation. By integrating LLMs into both the perception and action layers, A4FN illustrates how foundational models can drive autonomous control beyond traditional NLP applications.

Moreover, A4FN is also designed for seamless integration with current and future wireless systems, including 5G/6G and Wi-Fi 7/8. Its modular design supports deployment across a spectrum of environments -- from disaster zones to dense urban centers -- and can operate in both infrastructure extension and standalone modes.

To ensure interoperability, A4FN can interface with Software-Defined Networking (SDN) and Network Function Virtualization (NFV) frameworks via standard APIs. The DAA may interact with slicing orchestrators or 5G service-based components to enforce intent-derived policies across access and backhaul layers. Integration with Multi-access Edge Computing (MEC) platforms enables low-latency inference and localized autonomy. By combining programmable infrastructure with Agentic AI reasoning, A4FN lays the foundation for next-generation, self-synthesizing networks that are resilient, adaptive, and semantically aware.

\section{Challenges and Research Directions \label{sec:challenges}}
Realizing the A4FN vision requires addressing several research challenges across multimodal perception, resource-constrained inference, agent coordination, and scalable deployment. This section outlines six critical areas and associated research directions.

\subsection{Agent Instantiation and Deployment Models}
A core strength of A4FN lies in its architectural flexibility. However, determining optimal placement of the PA and DAA -- whether onboard UAVs, at the edge, or in the cloud -- requires careful analysis of trade-offs between latency, autonomy, compute resources, and energy efficiency. Deploying the PA on UAVs enables low-latency perception and localized intent inference but is constrained by onboard processing power and battery constraints. In contrast, hosting the DAA at edge/cloud locations supports centralized orchestration and richer decision-making but may introduce harmful communications delays in mission-critical scenarios. Hybrid configurations, where lightweight PA modules run on UAVs and the DAA resides at the edge, may provide a balanced solution.

Future research should explore modular agent architectures, dynamic task offloading, and migration strategies. Agents must scale functionality and relocate across infrastructure tiers in response to changing mission contexts, network conditions, and resource constraints.

\subsection{Multimodal Perception and Context Awareness}
The Perception Agent (PA) must fuse and interpret multimodal data -- including visual, acoustic, environmental, and network telemetry -- in real time and under adverse conditions. Challenges such as high dimensionality, sensor noise, occlusion, and data sparsity -- particularly in disaster scenarios -- complicate timely intent inference and increase fusion latency. Transformer-based architectures and graph neural networks offer promising solutions for modeling complex cross-modal relationships. However, additional techniques may be required to improve robustness and interpretability. One key direction is the development of salient signal prioritization mechanisms, capable of emphasizing contextually relevant features, such as thermal anomalies and human distress cues, while filtering redundant or irrelevant input. Another direction is the integration of adaptive attention frameworks that dynamically focus on the most informative modalities based on environmental context and mission urgency. Temporal memory mechanisms are also relevant. Differentiating between transient events (e.g., a passing siren) and persistent signals (e.g., ongoing fire, emergency activity) helps reduce false SLS triggers and supports more reliable, perception-driven control.

Together, these research directions can significantly enhance the PA’s ability to deliver accurate, context-aware intent inference in real-world environments.

\subsection{Semantic Reasoning and Intent Translation}
Central to A4FN is the paradigm of intent-driven networking, in which user needs and operational contexts are abstracted into SLSs. The generation and interpretation of these specifications depend on the PA’s semantic reasoning capabilities and the DAA’s ability to act upon inferred goals. Inaccuracies or inconsistencies in this process can lead to suboptimal resource allocation or degraded mission outcomes. Translating multimodal cues into actionable intents requires domain-specific grounding. Fine-tuning LLMs with datasets drawn from public safety operations, emergency communications, or tactical mission scenarios could enhance the accuracy and relevance of intent inference. Furthermore, implementing real-time feedback mechanisms -- where the DAA reports the outcome of reconfigurations in terms of achieved QoS -- can close the decision loop and enable the PA to refine future inferences. Coordination among multiple agents also demands a shared semantic context. 

Future research should explore semantic consistency frameworks that reconcile asynchronous, partial, or delayed updates across distributed PA and DAA instances, particularly in environments with intermittent or degraded connectivity.

\subsection{Inter-Agent Coordination and Synchronization}
Effective coordination between PA and DAA agents is crucial for consistent system behavior. Misaligned interpretations, stale data, or delayed intent propagation may lead to degraded performance or conflicting actions. 

Addressing this requires further research into decentralized and fault-tolerant communications protocols, machine learning-based agent synchronization, and asynchronous coordination frameworks. Such techniques will be critical to maintain system resilience in environments characterized by intermittent connectivity or degraded network conditions.

\subsection{Energy Efficiency and Resource-Constrained Inference}
Energy is a critical resource in UAV-based systems, influenced by both mobility and compute-intensive onboard tasks. Incorporating LLM-based agents onboard increases this demand, asking for energy-aware design at all system levels.

Model compression techniques (e.g., pruning, quantization, and distillation) are essential for deploying LLMs on UAVs or at the edge. Adaptive model scaling, where complexity adjusts based on energy availability and mission urgency, can further balance performance and power consumption. Decisions made by the DAA, particularly regarding UAV placement and reconfiguration, must also consider energy trade-offs. For example, repositioning UAVs to improve link quality must be balanced against the cost of propulsion. Further gains in sustainability can be achieved by integrating renewable energy sources (e.g., solar panels) and leveraging hardware accelerators (e.g., Tensor Processing Units) to reduce inference latency and power consumption.

\subsection{Scalability, Simulation, and Validation Environments}
Deploying A4FN in wide-area or dense-traffic scenarios presents challenges related to coordination, interference management, and system validation. Coordinating large swarms of UAVs requires scalable, fault-tolerant communications protocols -- particularly in non-line-of-sight or degraded environments.

Hierarchical or cluster-based control architectures can support scalable management, where localized decision-making is guided by global semantic intents. Coordination among distributed PA and DAA agents must be robust to communications latency and partial observability. Asynchronous agent synchronization and decentralized inference models are promising techniques for enabling real-time collaboration in fragmented or bandwidth-limited networks.

A major limitation is the lack of simulation platforms tailored to A4FN. Validating intent-driven control over multimodal inputs, LLM-based reasoning, and dynamic network topologies requires integrated environments. These should combine network simulators (e.g., ns-3), UAV flight dynamics (e.g., PX4, Gazebo), and AI toolchains (e.g., Hugging Face, ONNX Runtime). Realistic digital twins -- accounting for environmental variability, signal propagation, and UAV mobility -- are essential for testing, benchmarking, and policy learning.

\subsection{Cross-Cutting Challenges}
Beyond core functionality, explainability and trustworthiness of Agentic AI components are paramount, especially in safety-critical environments where transparency and auditability enable supervision and regulatory compliance. Security is another major concern, as LLM-based agents may be vulnerable to adversarial manipulation (e.g., malicious prompts, spoofed inputs). Defensive measures, including prompt validation, adversarial training, intent verification, and secure multi-agent protocols, are essential to ensure safe, reliable operation. As A4FN scales across distributed deployments, semantic negotiation and intent alignment among independently trained agents become increasingly important. Federated learning, consensus-based decision-making, and trust-aware reasoning can help preserve coherence, reliability, and resilience in multi-agent networks.

Addressing these challenges is key to evolving A4FN from concept to a scalable Agentic AI system. By embedding Agentic AI into FN control, A4FN aligns with emerging standards (e.g., 3GPP, ETSI MEC, IEEE UAV), supporting broader adoption. This paper lays the foundation for intelligent, adaptive telecom infrastructures capable of resilient performance in critical scenarios, reshaping wireless networks for a rapidly evolving digital society.

\section{Conclusions \label{sec:conclusions}}
This position paper presented A4FN, an Agentic AI conceptual architecture for the autonomous and intent-driven management of FNs. A4FN addresses the limitations of traditional FN approaches, namely static configurations and human-in-the-loop control, by introducing LLM-based agents capable of semantic reasoning and real-time reconfiguration. The architecture is built around two agentic components: the PA, which interprets multimodal sensor data to derive dynamic Service Level Specifications (SLSs), and the Decision-and-Action Agent (DAA), which translates these specifications into autonomous UAV placement, resource allocation, and network adaptation. By integrating semantic reasoning into the FN control loop, A4FN contributes to the realization of resilient, scalable, and self-managing network infrastructures. 

Future research will focus on instantiating the architecture in real-world scenarios, with particular attention to efficient model deployment, energy-aware operation, and robust multimodal perception. Ongoing efforts will also explore integration with emerging wireless technologies and distributed learning frameworks to enable broad applicability and sustainable deployment.

\bibliographystyle{IEEEtran}
\bibliography{references}

\end{document}